\documentclass[10pt, conference, letterpaper]{IEEEtran}

\ifCLASSINFOpdf

\else

\fi

\usepackage{flushend}
\usepackage{epsfig}
\usepackage{graphicx}
\usepackage{grffile}
\usepackage{longtable,ltxtable,booktabs}
\usepackage{forloop}
\usepackage{multirow}
\usepackage{acronym}
\usepackage{hyperref}
\usepackage{float}
\usepackage{enumitem}
\usepackage[dvipsnames]{xcolor}
\usepackage{enumitem}
\usepackage[labelfont=bf]{caption}
\setlist[itemize]{leftmargin=*}

\begin{document}

\title{Tracking the Big NAT across Europe and the U.S.}

\author{\IEEEauthorblockN{Anna Maria Mandalari}
\IEEEauthorblockA{University Carlos III of Madrid, Spain\\
}
\and
\IEEEauthorblockN{Andra Lutu}
\IEEEauthorblockA{Simula Research Laboratory, Norway\\
}
\and
\IEEEauthorblockN{Amogh Dhamdhere}
\IEEEauthorblockA{CAIDA/UC San Diego, CA\\\
} 
\and 
\hspace*{4cm}
\IEEEauthorblockN{Marcelo Bagnulo}
\IEEEauthorblockA{\hspace*{4cm}University Carlos III of Madrid, Spain\\
}
\and
\IEEEauthorblockN{KC Claffy}
\IEEEauthorblockA{CAIDA/UC San Diego, CA\\
}}

\maketitle

\begin{abstract}
Carrier Grade NAT (CGN) mechanisms enable ISPs to share a single IPv4 address across multiple customers, thus offering an immediate solution to the IPv4 address scarcity problem. 
%Though for mobile providers CGN technology has always been common, (at the time of writing) we have little confirmed data about the degree of CGN penetration in fixed-line broadband networks. 
In this paper, we perform a large scale active measurement campaign to detect CGNs in fixed broadband networks using NAT Revelio -- a tool we have developed and validated.
Revelio enables us to actively determine from within residential networks the type of upstream network address translation, namely NAT at the home gateway (customer-grade NAT) or NAT in the ISP (Carrier Grade NAT).
We demonstrate the generality of the methodology by deploying Revelio in the FCC Measuring Broadband America testbed operated by SamKnows and also in the RIPE Atlas testbed.
We enhance Revelio to actively discover from within any home network the type of upstream NAT configuration (i.e., simple home NAT or Carrier Grade NAT).
We ran an active large-scale measurement study of CGN usage from 5,121 measurement vantage points within over 60 different ISPs operating in Europe and the United States.
We found that 10\% of the ISPs we tested have some form of CGN deployment.
We validate our results with four ISPs at the IP level and, reported to the ground truth we collected, we conclude that Revelio was 100\% accurate in determining the upstream NAT configuration for all the corresponding lines. 
To the best of our knowledge, this represents the largest active measurement study of (confirmed) CGN deployments at the IP level in fixed broadband networks to date.
\end{abstract}

\IEEEpeerreviewmaketitle

\section{Introduction}

In light of the IPv4 address scarcity problem, one approach towards prolonging the life of current IPv4 address allocations is
to deploy Carrier Grade NATs (CGNs), where Internet Service Providers (ISPs) share the same public IPv4 address across multiple end users. 
CGNs may introduce a number of issues
for end users, service providers, content providers and government
authorities~\cite{rfc6269}. There is some evidence that CGNs can cause
dropped services in peer-to-peer applications, and lead to low
performance of file transfer and video streaming
sessions~\cite{rfc7021}. CGNs also introduce security challenges including
traceability of IP addresses and anti-spoofing. 
Despite these challenges, CGNs offer an immediate relief to the IPv4
address scarcity problem, so it is likely that their popularity will increase over time. 
The use case for CGN differs in wireline vs. mobile networks. 
Given the rapid boost in the number of operational Internet-enabled mobile devices, and the
scarcity of available IPv4 address space, mobile operators usually assign the same public IP address to multiple end users. 
Hence, some form of carrier-grade NAT technology has always been the norm, rather than the exception~\cite{Wang:2011:Untold, triukose2012geolocating}. 
In wireline networks, however, end users are typically assigned public IPv4 addresses. 
While this situation may change as IPv4 addresses become increasingly scarce and ISP customer
bases continue to grow, there is no source of systematic measurements of the prevalence or
evolution of CGN deployments in ISPs.

In this paper, we perform a large scale active measurement campaign to detect CGNs in fixed broadband networks using NAT Revelio~\cite{lutu2016nat}. 
Revelio enables us to actively determine from within residential networks the type of upstream network address translation, namely NAT at the home gateway (customer-grade NAT) or NAT in the ISP (Carrier Grade NAT). 
We deployed Revelio on two large-scale hardware-based measurement platforms -- RIPE Atlas in Europe and the FCC “Measuring Broadband America” (FCC-MBA) in the U.S. -- with a total of 5,121 vantage points in over 60 ISPs. 
The FCC-MBA deployment consisted of 2,477 home routers operated by SamKnows in 21 large residential broadband Internet access service providers in the U.S. 
We also adapted the Revelio methodology to run on the RIPE Atlas infrastructure (using their available user tests), and executed the tests from 2,644 Atlas probes in 43 ISPs mainly active in Europe. 
We thus demonstrate the flexibility of Revelio and exemplify the use of two fundamentally different large-scale measurement testbeds.

We ran the measurement campaign in two phases (May 2016 and August 2016) on both platforms. 
Based on the experimental results from the first phase (May 2016), we evolved the methodology of NAT Revelio to tackle a number of corner cases that confused our original approach. 
We enhanced the test suite to account for a wide diversity of home network topologies and various access technologies. 
In the second phase of the measurement campaign (August 2016) we deployed the evolved Revelio suite to investigate the state of CGN deployment in broadband networks.

Our contributions in this paper are twofold. 

First, we performed a large measurement campaign to detect CGN in fixed broadband access networks. 
We present the results of using (evolved) Revelio across 64 ISPs across Europe and the U.S. 
Our results show that 10\% (6 out of 64) of the ISPs we tested have some form of CGN deployment. 
In particular, one ISP has a large-scale deployment where Revelio detected upstream CGN deployment from all 76 vantage points in that ISP. 
In the other 5 ISPs we observed evidence of a localized deployment limited to a subset of customers. 
We verified our results with representatives of the ISPs to validate our positive and negative inferences at the IP level. 
We confirmed the results for 4 of the 6 positive ISPs by personal communications with ISP representatives. 
The combination of the FCC-MBA and RIPE Atlas study represents (to the best of our knowledge) the largest active measurement study to date with confirmed CGN deployments in broadband networks at the IP-level granularity.

Second, as a result of this large-scale measurement campaign, we added additional tests to Revelio to decrease the probability of false positives when quantifying the degree of CGN penetration in broadband providers. 
After closely analyzing the measurement dataset we collected in the first phase, we observed myriad cases of non-standard home network topologies and atypical results. 
We detail (Section~\ref{sec:evolve_revelio}) the two main setups that confused our original Revelio methodology, and the changes we made to the methodology to account for these corner cases. 
We have released the enhanced NAT Revelio test suite as an open source tool\footnote{NAT Revelio on github: \url{https://github.com/mami-project/revelio}} available to any user to download and run locally.

\section{NAT Revelio}
\label{sec:methodology}
In this section we describe NAT Revelio, the test suite we designed to actively detect NAT in the ISP access network. 
We presented an earlier version of NAT Revelio in~\cite{lutu2016nat}.
In particular, we highlight here the changes we bring to improve Revelio's accuracy by leveraging our extensive measurement campaigns in different testbeds.
We expand on the methodology in the context of residential Internet customers.
It is, however, possible to use the proposed methodology in other scenarios (e.g., corporate networks).

\begin{figure}[t]
	\centering
		    \includegraphics[width=0.5\textwidth]{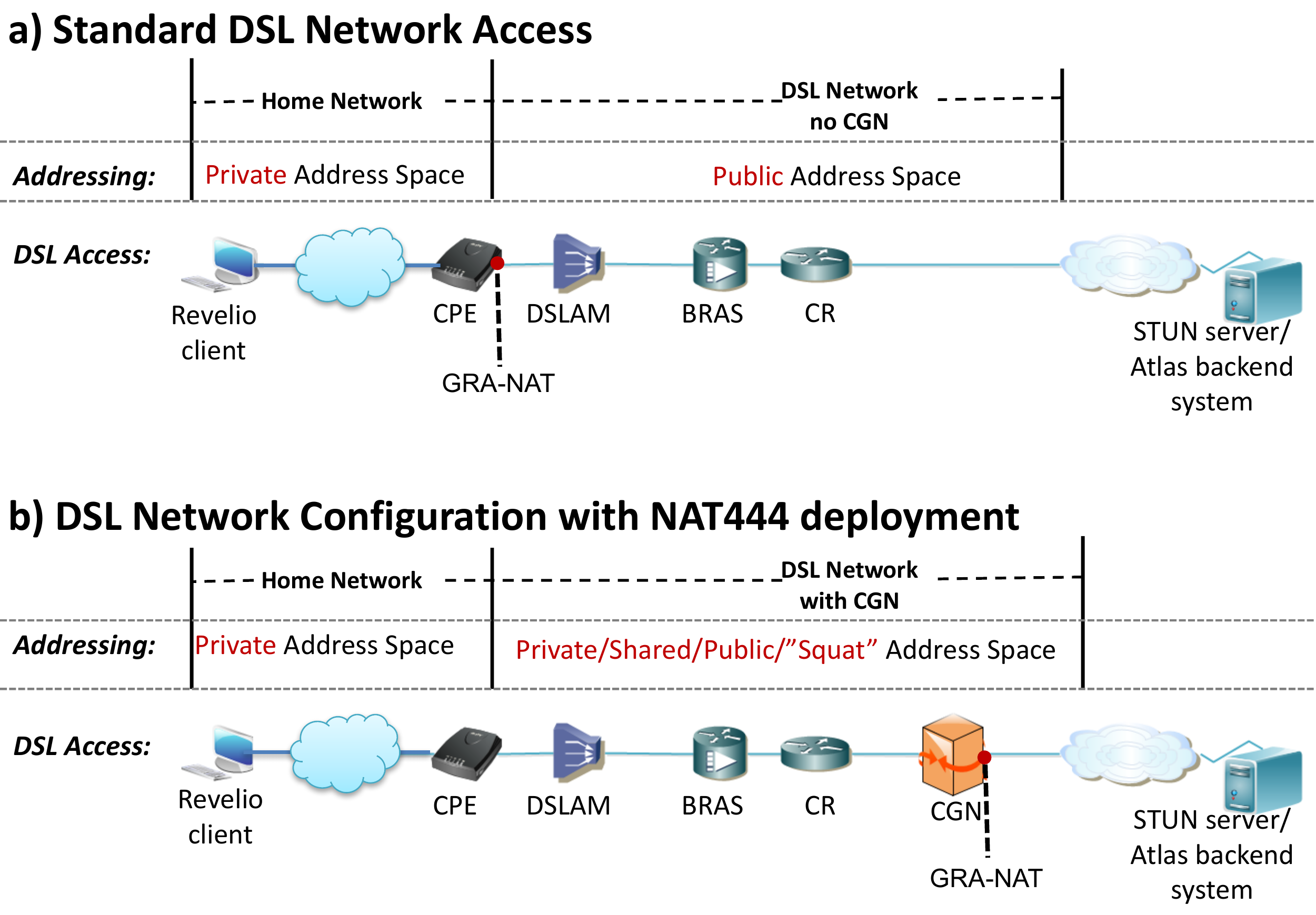}
	\caption{\em Revelio experimental setup for a DSL access network (the DSLAM, BRAS and CR are standard elements in the DSL architecture). The Revelio client runs on a device connected to the home network, whose exact topology we do not know. In the case of (a) standard DSL network access, the CPE performs the translation to the GRA, thus it is the GRA-NAT. In the case of (b) standard DSL network configuration with NAT444 deployment, the CGN is the one performing the translation to the GRA, thus it is the GRA-NAT. }
	\label{fig:setup}
	\vspace{-4mm}
\end{figure}

In Figure~\ref{fig:setup} we depict the residential setup we consider for Revelio in the context of a DSL access network. 
The home network may have an arbitrary topology consisting of multiple hosts, routers and switches including multiple levels of NATs. 
The home network connects to the Internet through the Customer Premises Equipment (CPE) also known as  ``home router" or  ``home gateway". 
The access link connects the CPE with the ISP access network. 
In the case of DSL technology, the access network includes the digital subscriber line access multiplexer (DSLAM), the broadband remote access server (BRAS) and the Core Router (CR).
The ISP network connects with the rest of the Internet. 

In terms of IP addressing, the home network generally uses private IP address space. 
The home gateway usually performs the NAT function (home-NAT) from the private addresses within the home network to the addresses used in the ISP access network which may be public, private or shared \cite{rfc6598}.
In some cases, end-users can configure several different realms of private addresses within their home network in the context of cascaded home NATs. 
Independently of the home network topology, when a host within the home network communicates with a host in the rest of the Internet, the private address used by the host in the home network translates to a public address that we call the Globally Routable Address (GRA). 
For the majority of the residential Internet market, the ISP configures the GRA on the Internet-facing interface of the CPE and the NAT function in the CPE translates from the private addresses in the home network to the GRA~\cite{garland2000communications}. 
An alternative, incipient, setup is one including an additional NAT function that operates in the ISP network (in addition to the NAT function in the CPE) 
and performs the final translation to the GRA. 
These configurations are usually called Carrier Grade NAT (CGN), Large Scale NAT (LSN) or NAT444. 
In this case, packets flowing between the home network and the Internet go through two upstream NAT-capable devices: the CPE (customer grade NAT) and the ISP NAT (Carrier-Grade NAT). 
The goal of the Revelio methodology is detect CGNs by distinguishing whether the NAT function translating to the GRA is located within the home network or it is located in the ISP network. 

In order to discern where the translation to the GRA occurs, Revelio performs active tests from a device connected to the home network. 
The \emph{probe} running Revelio connects to the home network and may or may not be directly connected to the CPE, i.e., there may be multiple hops, including ones performing NAT function(s), between the \textit{probe} and the CPE. 
The target of the active tests the \emph{probe} performs are servers located in the Internet (Figure~\ref{fig:setup}).
NAT Revelio does not require any cooperation from the ISP beyond forwarding Internet packets to and from the customer. 

\subsection{Revelio Methodology Overview}

As mentioned earlier, the purpose of Revelio is to detect whether the \textit{device performing the translation to the GRA} (hereinafter, the GRA-NAT) \textit{resides in the home network or in the ISP network}. 
In order to do this, Revelio attempts to pinpoint the location of the GRA-NAT with respect to the access link. 
If the GRA-NAT lies between the \textit{probe} and the CPE, we conclude that the user is not behind a CGN. 
If the GRA-NAT lies after the CPE, we conclude that the ISP deploys CGN.  
In order to achieve this, Revelio needs to determine the location of the GRA-NAT and the location of the access link with respect to the \textit{probe} and compare them. 

\paragraph{Initial Revelio tests}
To determine the location of the GRA-NAT, Revelio performs the following steps: 
1) discovers the GRA by running STUN~\cite{rfc5389} against a public STUN Server located in the Internet (FCC MBA) or by using the Atlas API (RIPE Atlas);
2) runs traceroute to the GRA, computing the number of hops from the probe to the GRA-NAT;

The determination of the location of the access link is challenging because we aim to support arbitrary topologies in the home network and we do not have any prior information about where in the home network the \emph{probe} connects. 
To determine the location of the access link we make the following assumption: the propagation delay of the access link is one order of magnitude or higher than the propagation delay of the links in the home network. 
We believe this is a realistic assumption for the different access technologies and home network technologies available in the market and it is supported by existing empirical evidence~\cite{sundaresan2011broadband}. 
In order to locate the access link, we estimate the propagation delay of different links between the \textit{probe} and an arbitrary target server in the Internet using \emph{pathchar}~\cite{pathchar_study}. 
\emph{Pathchar} is a well-known technique for estimating transmission and propagation delays along the path using multiple traceroute measurements with different packet sizes. 
Since we only need to estimate the order of magnitude of the delays, the precision of \emph{pathchar} is sufficient. 
Using the propagation delays measured by \emph{pathchar}, we determine the access link as the first one with a propagation delay at least one order of magnitude higher than the previous links. 

By comparing the respective locations obtained for the GRA-NAT and for the access link, we can establish whether the GRA-NAT is located \textit{before} the access link (no CGN) or \textit{after} the access link (ISP uses CGN).

\paragraph{Supplementary Revelio tests}
In addition to the main detection tests, the Revelio client performs two other tests to gain additional insight about the presence of CGNs. 
In particular, Revelio performs the following tests:

%{\bf UPnP actions:} If we determine that the \textit{probe} directly connects to the CPE (i.e. the access link is one hop away), Revelio tries to run the UPnP protocol~\cite{upnp} from the \emph{probe} to query the CPE for the IP address allocated to its external interface. 
%If this matches the GRA, then we can conclude that there is no CGN.
%Otherwise, if the UPnP query returns a private/shared address, we can conclude there is an upstream CGN.
%{\bf Private/shared addresses along the path:} We search for private and shared addresses the ISP configures on network devices after the access link. 
%The presence of private/shared addresses by itself does not imply the presence or absence of a CGN, but it serves as a hint that this may be the case. 
%In particular, the presence of shared addresses provides a stronger hint about the presence of CGNs, as this address block is specifically destined for this purpose~\cite{rfc6598}.

{\bf Invoke UPnP actions.} If the \textit{probe} directly connects to the CPE (i.e., the access link is one hop away), Revelio tries to run the UPnP protocol~\cite{upnp} from the \emph{probe} to retrieve the IP address of the WAN-facing interface of the CPE. 
If this IP address matches the GRA, we conclude that there is no CGN.
Otherwise, if the UPnP query returns a private/shared address, Revelio detects an upstream CGN.

{\bf Private/shared addresses along the path:} To detect the access link location, Revelio runs multiple traceroute measurements to a fixed target. 
This enables us to retrieve the IP addresses operators configure in their network.
We then search for private and shared addresses after the access link. 
The detection of private/shared addresses after the access link alone does not imply the presence of an upstream CGN, but it serves as a hint that the ISP might be operating one. 
In particular, the presence of shared addresses after the access link provides a stronger indication about the presence of CGNs, because this address block is specifically reserved for CGN deployment~\cite{rfc6598}.

\subsection{Evolving Revelio}
\label{sec:evolve_revelio}

After the first phase of the large scale measurement campaign (May 2016) with Revelio on RIPE Atlas and FCC MBA and communicating with several of the ISPs we measure, we identified some corner cases that may confuse Revelio, which we describe in detail below.
The common link between these corner cases is the fact that they counter the logic of the access link detection approach we integrated in the original Revelio test-suite.
Given Revelio's reliance on the correct detection of the access link location, incorrectly mapping non-standard home network topologies leads to false positives.
To tackle these particular issues and increase the robustness of Revelio, we enhance the original methodology by adding the following two tests to the test-suite.

{\bf Pathchar to the GRA.} We identified 165 home gateways that replied to traceroute to the GRA as if they were two hops (148 in RIPE Atlas and 17 in FCC MBA), thus generating spurious links in the results.
When processing the traceroute to the GRA from the \textit{probe}, the home gateways generate one reply from the internal interface (for packets with TTL=n) and a second reply from the external WAN-facing interface that assigns the GRA (for packets with TTL=n+1). 
When running pathchar to the external server, however, the home gateway only replies to traceroute from the internal interface (for packets with TTL=n), while the subsequent reply comes from the following hop (for packets with TTL=n+1).
This behavior breaks the Revelio methodology because the GRA appears to be one hop farther from the \textit{probe} than it actually is, making us believe that the GRA-NAT is past the access link.  
We detected this behaviour in several models of home routers, including SpeedPort or FritzBox.

In order to control for these cases, we run \textit{pathchar} from the \textit{probe} to the GRA, similarly to the pathchar to the external server.
The home gateway replies as if it were two hops, generating a spurious link that masquerades as the access link. 
We contrast the access link propagation delay value we obtained running \textit{pathchar} to the external server with the one we obtained running \textit{pathchar} to the GRA and observe that they are significantly different.  
If the CPE generates two replies to the traceroute to the GRA, the delay of the spurious link from the CPE to the GRA measured by the \textit{pathchar} to the GRA is significantly smaller than the delay of the \textit{real} access link measured by the \textit{pathchar} to the external server. 
This is so because the spurious link is internal to the CPE, while the link we measure with \textit{pathchar} to the external server is the actual access link.  
By comparing the two \textit{pathchar} results, we can identify these anomalous CPEs and correctly identify the GRA at hop n.

\begin{figure}[t]
	\centering
		    \includegraphics[width=0.5\textwidth]{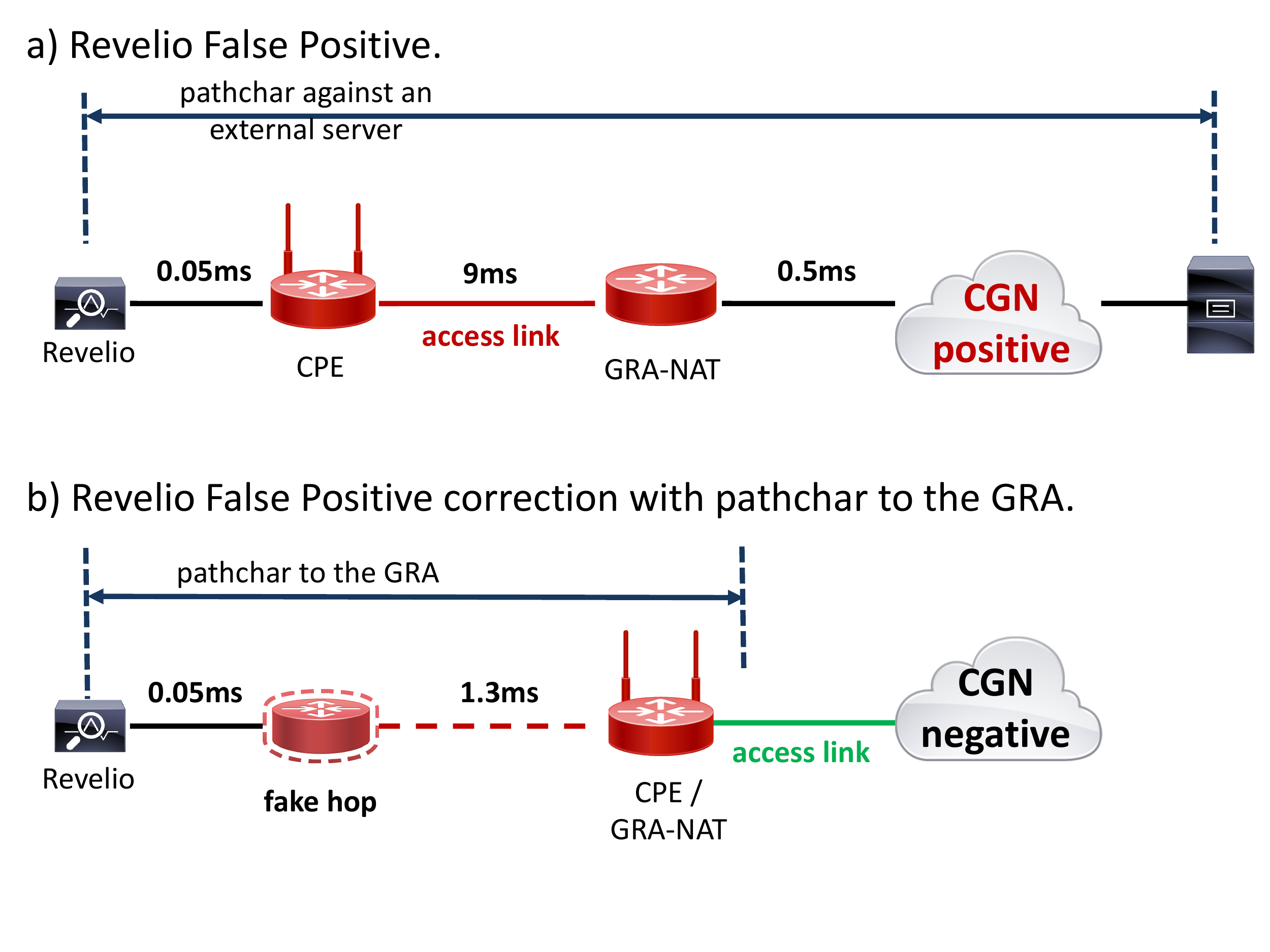}
	\caption{\em We observed that some home gateways reply to traceroute as if they are two hosts, generating spurious links. When running pathchar to an external server, Revelio shows that the access link is the spurious link (Figure (a)). We run pathchar to the GRA (Figure (b)) and compare the results with the previous output from pathchar to the external server to correct these Revelio false positives. }
	\label{fig:spurious_links}
	\vspace{-4mm}
\end{figure}

We exemplify this non-standard behavior in Figure~\ref{fig:spurious_links}.
The original Revelio suite generated positive results for upstream CGN because the GRA-NAT appears to be outside the home network (Figure~\ref{fig:spurious_links}(a)) based on the \textit{pathchar} to the external server test and the number of hops until the GRA (from traceroute to the GRA). 
The \textit{pathchar} to the external server results shows that the access link is the red link with a delay of 9ms.
In the same time, we see that the GRA-NAT is after the access link, in the access network, based on the results of the traceroute to the GRA.
However, when running \textit{pathchar} to the GRA, we obtain a different value for the propagation delay on what seems to be access link (Figure~\ref{fig:spurious_links}(b))).  
The dotted red link is actually not the access link, but the spurious link that is internal to the CPE.
Given the large difference between the value of the propagation delay on the spurious link (Figure~\ref{fig:spurious_links}(b))) and the propagation delay on the actual access link (Figure~\ref{fig:spurious_links}(a)), we conclude that the home router replied to the traceroute to the GRA as two hops and Revelio generated a false positive. 

 {\bf Expected access technology delay.} 
 In some cases, we have detected that the propagation delay of the different links within the home network differs in one order of magnitude (e.g., one link with a delay of tens of $\mu$s and another one with delay in the hundreds of $\mu$s), confusing the Revelio methodology. 
In this case, the delays of both home network links are still one order of magnitude less than the propagation delay of the actual access link. 
In order to deal with this case, we define an expected range for the access link delay based on the access technology and we verify if the access link delay we measure falls within the expected range. 
If this is not the case, we mark the first link that falls within the expected range as the access link. 
In the case of the FCC MBA platform, we know the access link from the per-probe metadata we received.
For RIPE Atlas, we leverage the user tags that also contain information about the access technology. 
 
\begin{figure}[t]
	\centering
		    \includegraphics[width=0.5\textwidth]{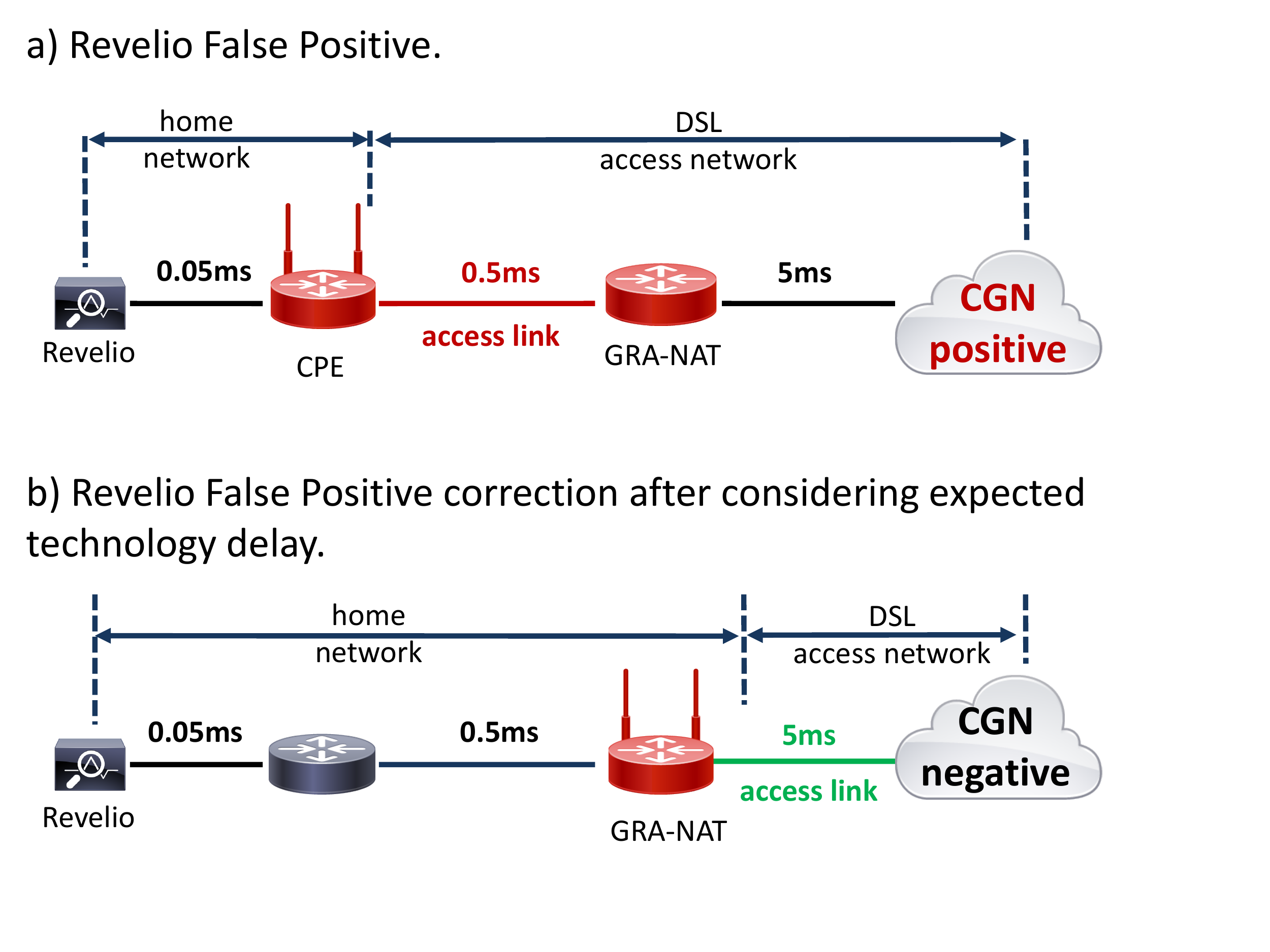}
	\caption{\em We observed that propagation delay of the different links within the home network differs in one order of magnitude. When applying the original Revelio methodology, we erroneously conclude that the GRA-NAT is in access network of the ISP (Figure (a)). When factoring the expected propagation delay corresponding to the access technology we correct the Revelio false positive (Figure (b)). }
	\label{fig:technology_delay}
	\vspace{-4mm}
\end{figure} 
 
\begin{figure}[t]
	\centering
		    \includegraphics[width=0.5\textwidth]{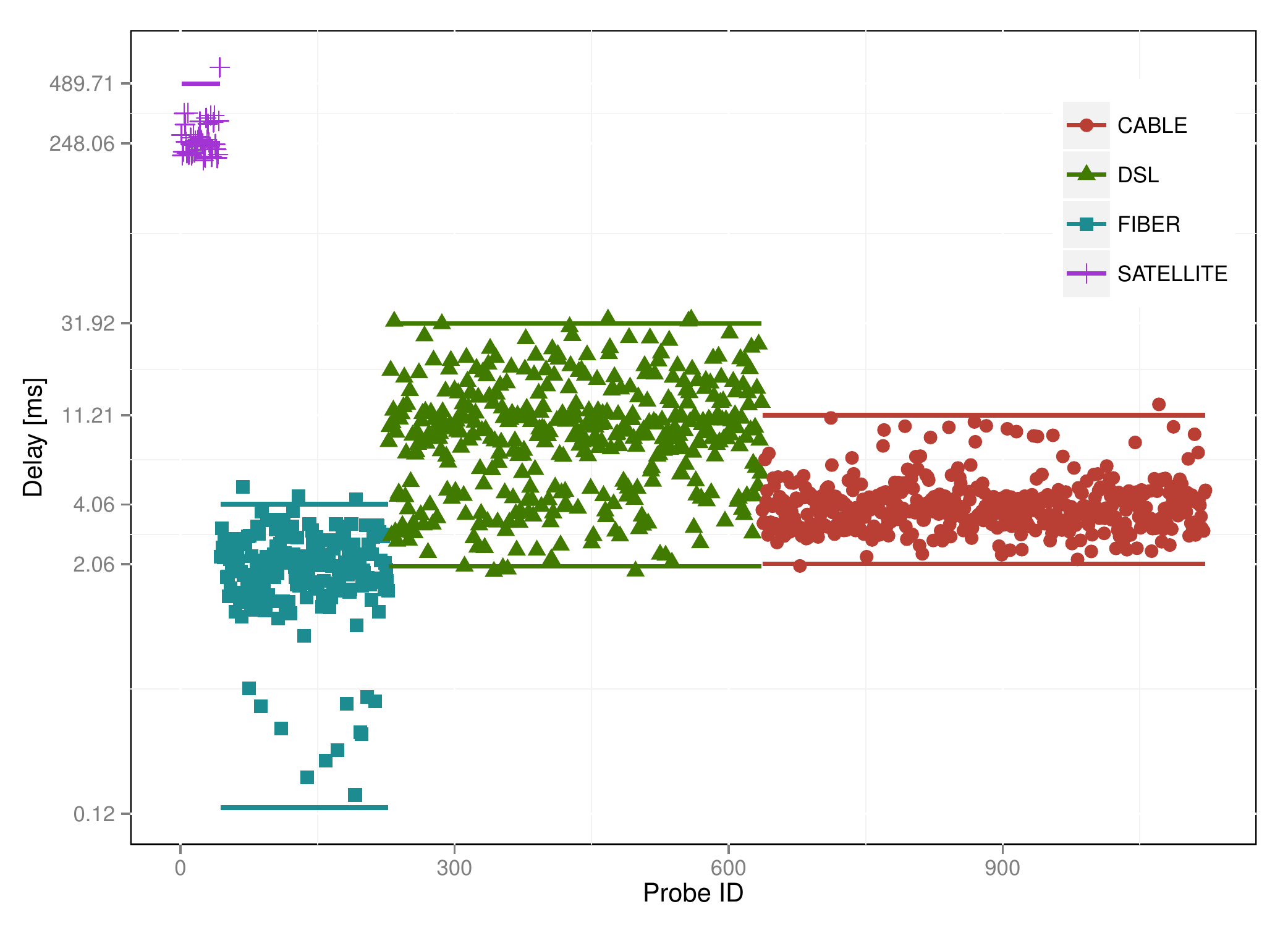}
	\caption{\em Scatter plot of the access link delays we identified using the original Revelio methodology. The horizontal lines represent the range od expected access link propagation delay per access technology. The points that lie outside the expected range are potential false positives in the Revelio results. }
	\label{fig:scatter_delay}
	\vspace{-4mm}
\end{figure}

We exemplify such an example in Figure~\ref{fig:technology_delay}.
We first observe that the two red routers in Figure~\ref{fig:technology_delay}.(a) reply to the traceroute probing with private addresses. Then, after running pathchar we obtain the propagation delay for the links.  
The first link with a propagation delay higher by one order of magnitude than the previous one is the one we mark as the access link. 
Because we detect that the GRA-NAT is after the access link (and we also detect private addresses in the access network) we conclude that there is an upstream CGN active. 
However, when comparing the propagation delay on the access link with the expected propagation delay corresponding to DSL access (i.e., between 2ms and 30ms), we conclude that the first two links are part of the home network, and the access link is the green link in Figure~\ref{fig:technology_delay}.(b).
In Figure~\ref{fig:scatter_delay} we show the spread of access link propagation delay values we obtained using the original Revelio methodology. 
The horizontal lines show the expected range of propagation delay per access technology. 
The points that fall outside the expected range are potential false positives which we tackle within the evolved Revelio suite.

%\paragraph{Revelio validation}
%With the help of a large UK ISP and a large Italian ISP, we tested NAT Revelio on a controlled set of 30 operational DSL lines, 2 of which connected behind a trial CGN implementation in the same ISP.  
%NAT Revelio accurately detected the upstream NAT configuration of all 30 lines.  

\subsection{Revelio Limitations}
\label{sec:limitations}

The evolved Revelio methodology still suffers from a series of limitations, which we discuss in this section.

The Revelio methodology heavily relies on traceroute. 
If traceroute packets are filtered, the Revelio tests will be inconclusive because we cannot reliably locate the access link. 
%However, in the case of the traceroute to the GRA, if after a certain hop the traceroute packets are filtered, it is still possible to ascertain a lower bound for the distance to the GRA, which in turn can be compared to the number of hops to the access link, which may be enough to determine the CGN. 
However, in the case of traceroute to the GRA, receiving ICMP error replies from at least one hop immediately after the access link (i.e., after the CPE in Figure~\ref{fig:setup}) is enough to establish whether the GRA-NAT lies after the access link, in the ISP network.
In other words, even if the traceroute does not reach the target GRA, the Revelio methodology only requires for the packet to be routed outside the home network, in the ISP network.

Anecdotal evidence exists about home routers rate-limiting generation of ICMP packets~\cite{privatecommunication}, which can hinder the \emph{pathchar} methodology we use to measure the link propagation delays. 
Rate-limited ICMP errors will add additional delay to the traceroute messages, thus artificially increasing the delay we estimate for that link. 
Since we subtract ICMP round-trip delays to infer per-link delay, ICMP rate-limiting will mislead our inference.
%Following \emph{pathchar}'s assumption that the delay metric is additive, we subtract the delay added to this particular link from the subsequent link(s) in order to quantify the delay per link.
%These type of behavior then cannot be detected by searching for several links with high delay, as we do in the normal case. 
%If this occurs, the results may be false. 
Also, the lack of information about the type of access technology may increase the number of false positives in the case of non-standard home topologies. 
However, this is a non-issue when the end-user deploys Revelio to verify her upstream configuration. 
In the case of deploying Revelio on other hardware platforms or crowdsourcing the Revelio measurements, we assume this type of information is not hard to obtain from the platform operators. 

One of the Revelio tests requires the \textit{probe} to send UPnP queries to the CPE, which should also support UPnP. 
These devices do not always support UPnP, limiting the effectiveness of Revelio.
The Atlas probes do not allow users to perform UPnP queries. 
Additionally, for some of the SK probes, the CPEs do not support UPnP and cannot reply to the query the \textit{probe} sends. 
% Similarly, Revelio relies on UPnP and several \textit{probes} (all the Atlas ones) and many CPEs do not support UPnP, limiting the effectiveness of the test. 

\section{Revelio Deployments and Data}
\label{sec:measurements}

We deployed the evolved Revelio test-suite in two large measurement platforms, the FCC-MBA platform and RIPE Atlas, in August 2016. 
We next detail the two platforms and describe the dataset we obtained. 

\textbf{FCC-MBA platform:} The United States Federal Communication Commission's (FCC) initiative ``Measuring Broadband America" program~\cite{MBA-URL} is an ongoing nationwide performance study of broadband service in the United States that aims to improve the availability of information for consumers about their broadband service.
The initiative manages a large hardware-based measurement platform 
operated by SamKnows (SK), an international statistics and analytics firm supporting similar projects in other countries around the world.
The \textit{SK probes} are off-the-shelf routers that users voluntarily host in their home networks. 
The \textit{SK probes} run pre-installed software that measures Internet connection performance metrics including download speed, upload speed or latency.
Apart from the standard pre-installed measurements, the \textit{SK probes} can execute other custom tests.
We deployed the NAT Revelio client to run on 2,477 \textit{SK probes} operating as part of the MBA testbed.
This allowed us the unique convenience of running \textit{active measurements from inside the end-users home networks} across the U.S. and attempt to reveal the presence on a NAT in their ISP. 
%The recommended deployment configuration for the \textit{SK probes} is to connect it directly to the home gateway.
%Users often disregard this recommendation and deploy the \textit{SK probe} behind several cascaded in-home devices.
%Revelio can cope with such non-standard deployments because we designed it to support arbitrary home network topologies. 
For each probe in the FCC-MBA platform, we also have access to metadata regarding the access technology and ISP of each user line, which SamKnows collects and maintains. 
Based on this information, the FCC-MBA panel covers 23 different ISPs and tests 4 different access technologies overall.
These include 10 major DSL providers, 8 different cable providers, 3 ISPs that offer fiber to the home Internet connectivity and 2 main satellite providers in the U.S. 
In terms of aggregated number of \textit{probes} per technology, according to the metadata we have available, cable Internet access is predominant with approximately 40\% of the \textit{probes} operating on cable access.
DSL follows with a share of approximately 20\% of the \textit{probes} we tested.
Fibre access lines account for 9\% of the US-Revelio testbed.
 3.5\% of the testbed is made up of satellite access lines.

\textbf{RIPE Atlas platform:}
RIPE Atlas~\cite{atlas-url} is a measurement platform deployed and maintained by the RIPE NCC, consisting of thousands of hardware probes that perform various active measurements.
The Atlas platform offers a set of standard measurements as a service to its users, including ping and traceroute.
For the RIPE Atlas probes, we use the API to check the user tags regarding the access technology, which we retrieve from the information that the hosts of the probes volunteer to RIPE. 
At the time we deployed the Revelio suite, Atlas consisted of 9,231 active probes distributed in 3,381 Autonomous Systems (ASes), according to the RIPE Atlas API. 

To ensure a statistically significant sample of vantage points in an ISP, we only tested the ISPs with at least 20 Atlas probes. 
We aggregated the probes per AS using the probe metadata.
Thus, we deployed Revelio in 2,644 \textit{Atlas probes}, corresponding to 43 different ASes.
However, the set of tests that we were able to run in the \textit{Atlas probes} is more limited than the ones we can run on the \textit{SK probes}. 
In particular, the \textit{Atlas probes} do not provide support for UPnP.
Thus, as opposed to FCC-MBA where we ran the complete Revelio test suite, on the Atlas infrastructure we only deployed the traceroute-based tests.

\subsection{Revelio Dataset}

We scheduled the Revelio client to run over 20 times on each \textit{probe} during August 2016 for the FCC-MBA platform and in the RIPE Atlas platform.
The data we collected from the \emph{probes} in \emph{each} run of Revelio are the following:
the Globally Routable Address (GRA), the mapped port number, traceroute results to the GRA, traceroute results to a fixed target address (with 21 different packet sizes), 
%%@@AD: 19 times with 21 different packet sizes?
UPnP query result to retrieve the IP address on the external interface of the device to which the 
%2,477 
\textit{probes} connect (only for \textit{SK probes}).
In total, we collected data from 5,121 \textit{probes} in 64 ISPs with an average of 20 repetitions\footnote[2]{In the FCC-MBA platform, in order not to interfere with normal user Internet activity, the \textit{probes} perform cross-traffic sensing and run the tests we schedule only when they detect no end-user traffic. Thus, the number of Revelio repetitions differs for various measurement vantage points.}
per \textit{probe} which resulted in over 2 million traceroutes. 
The ISPs we tested include 42 DSL providers, 16 cable providers, 4 ISPs that offer fiber to the home Internet connectivity and 2 satellite providers. 
We are in the process of releasing the FCC-MBA Revelio dataset after proper anonymization. The RIPE Atlas measurements are publicly available and retrievable using the RIPE Atlas API. We can make available the measurement identifiers (MIDs) corresponding to Revelio tests upon request (interested parties can retrieve them to then query the RIPE Atlas database).

We processed the raw data we collect using Revelio and combined it with metadata we received from SamKnows and RIPE to build the \emph{Revelio State} for each \textit{probe}.
The Revelio State consists of the device identifier, ISP name, access technology, local IP address of the \textit{probe}, GRA, number of hops we measure until the GRA, location of the access link, set of private IPs we detect \emph{immediately} after the CPE (if any), set of shared IP addresses we detect after the access link (if any) and number of times Revelio successfully ran on the \textit{probe}. 
Revelio then feeds this information to the algorithm for detecting the type of upstream NAT.  
%%@@AD: perhaps we should say that the quality of the metadata might be different in the two cases: for FCC-Samknows I think it is more reliable, they do some verification. For ATlas it is mostly just user-volunteered. 

\section{Measurement Results}
\label{sec:results}

In this section, we present our analysis and validation of the NAT Revelio results we collected in August 2016. 

\subsection{Catching the Big NAT}
%%@@AD: too much italics in this section
We deploy NAT Revelio to detect the presence of CGN on a total of 5,121 different customer lines. 
In function of the upstream NAT configuration, Revelio classifies each \textit{probe} into one of the following cases: 
(i) \emph{inconclusive} (cases Revelio was unable to draw any conclusion due to incomplete or inconsistent results).
(ii) \emph{no home NAT} (i.e., the \textit{probe} where Revelio runs is directly connected to the public Internet), 
(iii) \emph{simple home NAT} (the CPE performs the GRA-NAT), 
(iv) \emph{Carrier Grade NAT} (the GRA-NAT is outside the home network, in the ISP's network) and 
We aggregate the results by the inferred upstream NAT configuration (Table~\ref{table:revelio_results}). 

%Moreover, we account for the limitations of the dataset we collect and associate to each detection result a confidence level that shows how likely it is that Revelio reports false positives. 

{\bf Inconclusive.}
For 1,276 \textit{probes} (307 \textit{SK probes} and 969 \textit{Atlas probes}), Revelio gave inconclusive results either because none of the tests could run on the \textit{probe} or because we did not obtain enough information to properly interpret the results we were able to collect. Our approach is conservative and tags as inconclusive the case of mixed responses from different tests. 
For example, traceroute limitations and ICMP traffic being filtered along the path to the external target server hamper our capacity to identify the access link (Section~\ref{sec:limitations}).
Without knowing the location of the access link, when the end-user deploys several levels of NAT in the home, we cannot draw conclusions regarding the presence of NAT in the ISP.
These \textit{probes} account for approximately 24\% of the total, (12\% of the \textit{SK probes} and 36\% of the \textit{Atlas probes}). 
We discard these cases from further analysis.

{\bf No home NAT. }
Revelio found that in 299 different cases (85 in \textit{SK probes} and 214 \textit{Atlas probes}), the Revelio client was running on a \textit{probe} configured with a public IP address that was also the GRA. 
These \textit{probes} were operating in the public Internet, which implies that the lines were not connected behind a NAT solution.
In all these cases, the \textit{traceroute to the GRA} test also confirmed the lack of a NAT solution in the corresponding ISPs. 

{\bf Simple home NAT. }
Out of the rest, for 3,454 \textit{probes} (2,009 \textit{SK probes} and 1,445 \textit{Atlas probes}) Revelio established the presence of simple home NAT and excluded the possibility of further NAT in the ISP.
Revelio reports the simple home NAT configuration (and, thus, the lack of NAT in the ISP for the respective line) when at least one of the \textit{traceroute to GRA} and \textit{invoking UPnP actions} tests establish that the home gateway is performing the GRA-NAT.
In the case of the UPnP test, for 1,300 \textit{SK probes} the address retrieved through UPnP from the CPE matched the GRA, concluding that the CPE was the GRA-NAT.  
For 815 \textit{SK probes}, the Revelio client was unable to communicate with the CPE through UPnP, either because the CPE did not supported UPnP or because the \textit{SK probe} was not directly connected to the CPE.
In the case of the traceroute to the GRA test, for 2,965 \textit{probes} (1,520 \textit{SK probes} and 1,445 \textit{Atlas probes}) Revelio located the GRA-NAT before the access link, concluding that the CPE was also the GRA-NAT. 
As a interesting data point, using \textit{pathchar to the GRA} test Revelio purged 165 of cases where the CPE replied as being two different hops, creating false positives.
In particular, Revelio detected this behavior in one single ISP for 78 out of 228 \textit{probes}. 

\renewcommand{\arraystretch}{2}
\begin{table*} [tbp!]
\centering
\scriptsize
\begin{tabular}{|c|c|c|c|c|c|c|c|}
\hline
 \multicolumn{1}{|p{1.8cm}|}{\centering \textbf{ISP ID}} & \multicolumn{1}{p{1.8cm}|}{\centering \textbf{CC}} & \multicolumn{1}{p{1.8cm}|}{\centering \textbf{Tech.}} &  \multicolumn{1}{p{1.8cm}|}{\centering \textbf{\# of probes}}& \multicolumn{1}{p{1.8cm}|}{\centering \textbf{ Inconclusive}} &\multicolumn{1}{p{1.8cm}|}{\centering \textbf{Simple Home NAT}} &\multicolumn{1}{p{1.8cm}|}{\centering \textbf{Carrier Grade NAT}} &\multicolumn{1}{p{1.8cm}|}{\centering \textbf{ Confirmed}} \\
\hline
\textbf{1 (Undisclosed ISP)} &US& Satellite & 76 & 0 & 0 & 76 & Yes\\ %Hughes
\hline
\textbf{2 (Kabel Deutschland)} &DE& Cable & 49 & 27 &14& 8 & Partially\\ %Kabel Deutschland
\hline
\textbf{3 (Fastweb)} &IT& Fiber & 26 & 14 & 8 & 4 & Yes \\ %FastWeb
\hline
\textbf{4 (OTE)} &GR& DSL & 21 & 5 & 14 & 2 & No Reply\\ %OTE
\hline
\textbf{5 (Liberty Global)} &NL& Cable & 280 & 133 & 146 & 1 & Yes \\ %Liberty Global
\hline 
\textbf{6 (Zen)} &UK& DSL & 32 & 11& 20 & 1 & No Reply\\ %Zen
\hline
%\hline
%\textbf{1} &US& Sat & 76 & 0 & 0 & 76 & Yes\\ %Hughes
%\hline
%\textbf{2} &US& DSL & 433 & 135 & 288 & 10 & No Reply\\ %AT&T
%\hline
%\textbf{3} &DE& Cable & 49 & 27 &14& 8 & Partially\\ %Kabel Deutschland
%\hline
%\textbf{4} &US& DSL & 92 & 1& 86 & 5 & No Reply\\ %Windstream
%\hline
%\textbf{5} &IT& Fiber & 26 & 14 & 8 & 4 & Yes \\ %FastWeb
%\hline
%\textbf{6} &GR& DSL & 21 & 5 & 14 & 2 & No Reply\\ %OTE
%\hline
%\textbf{7} &US& Cable & 759 & 142 & 615 & 2 & No Reply\\ %Comcast
%\hline
%\textbf{8} &NL& DSL & 280 & 133 & 146 & 1 & Yes \\ %Liberty Global
%\hline 
%\textbf{9} &UK& DSL & 32 & 11& 20 & 1 & No Reply\\ %Zen
%\hline
\end{tabular}
\vspace{0.1cm}
\caption{\small \textbf{Revelio Positive Results:} List of ISPs with at least one probe with \textbf{\emph{positive}} Revelio result (i.e., operates behind a CGN). 
We report the Country Code (CC), the access technology (Tech.), the total number of probes we tested for that ISP (\# of probes), the number of probes for which Revelio gave inconclusive results (Inconclusive), the number of probes Revelio tested negative (Simple Home NAT), the number of probes Revelio tested as positive (Carrier Grade NAT) and the current status of the confirmation from representatives of the ISP with positive Revelio results (Confirmed). For the latter, we mark this field with \emph{Yes} if the ISP confirmed the Revelio results at the IP level, \emph{Partially} if the ISP confirmed they use CGN but did not confirm the specific IP lines tested, \emph{No Reply} if we did not get any feedback from the ISP.}
\vspace{-6mm}
\label{table:revelio_results}
\end{table*}

{\bf Carrier Grade NAT. }
For 92 \textit{probes} in 6 ISPs (76 \textit{SK probes} in 1 ISP and 16 \textit{Atlas probes} in 5 ISPs) Revelio detected the presence of CGN technology in the ISP's network. 
Table~\ref{table:revelio_results} details the number of \textit{probes} that tested positive for CGN per ISP\footnote[3]{ We only disclose the names of the ISPs we tested using the RIPE Atlas platform. We are currently pending the approval of the FCC for disclosing the names of the ISPs we tested with the FCC-MBA testbed.}
%We include information about the total number of \textit{probes} in each ISP. 
We identified one satellite provider in the U.S. where all \textit{probes} tested positive for CGN.
%\footnote{For this ISP, Revelio reported an additional line with simple NAT upstream configuration (no CGN). However, we discarded this results after checking that the probe was mislabeled in the FCC-MBA testbed as belonging to ISP 5 (the end users connected, in fact, to a different provider). }. 
For the rest of the ISPs, we detected a mix of some \textit{probes} that tested positive for CGN and others that did not.
Overall, about 2\% of the \textit{probes} tested positive for CGN.
About 10\% of the ISPs we tested hosted at least one \textit{probe} that tested positive for CGN. 
Of these latter ones, only one ISP had a widespread deployment of CGN, while the other ISPs presented a few scattered \textit{probes} that tested positive, hinting a localized deployment, e.g., possibly for trials or suggesting a specific service.

\subsection{Validation of Revelio Results}

NAT Revelio tested 5,121 Internet lines in 64 different ISPs worldwide. 
In total, it reported 92 end users with an upstream CGN, which connected to 6 different ISPs. 
We validated both the positive (upstream CGN) and negative (no upstream CGN) results at the IP level through different means, including direct contacts with the involved ISPs or, in one case, using the WHOIS database information. 

\textbf{Positive Revelio Results.}
We obtained confirmations at the IP level from 4 ISPs (89 {\em probes}) for the presence of CGN in their network for the lines we tested and received no replies from the other 2 ISPs (3 {\em probes}). 
In Table~\ref{table:revelio_results} we report on the status on communication with the ISPs for which Revelio identified the presence of CGN. 
In particular, for ISP\#1 from Table~\ref{table:revelio_results} -- the satellite provider in the US for which all \textit{probes} tested positive -- the operator confirmed that its normal configuration includes performing the NAT function in the ISP network and that all the 76 lines that tested positive were indeed behind a CGN.
ISP\#3 (Fastweb) confirmed both the positive and the negative Revelio results.
For ISP\#5 (Liberty Global) from Table~\ref{table:revelio_results}, the GRA associated with the \textit{probe} is actually tagged in the WHOIS database (in the {\em Organization} field) as CGNAT (the other 279 \textit{probes} in the same ISP did not have a GRA in the subnet marked as CGN). 
ISP\#2 (Kabel Deutchland) from Table~\ref{table:revelio_results} confirmed that it is using CGN in its network. However, we did not obtain explicit confirmation from their representatives that the exact lines we detected as positive are actually behind a CGN, which is why we marked it as a \textit{partial} confirmation.

Based on the ground truth we collected, we conclude that NAT Revelio did not generate any false positives. 
Thus, provided that NAT Revelio can successfully run (see Section~\ref{sec:limitations} for limitations), its precision\footnote[4]{The precision represents the ratio between the number of true positives and the sum of the true positives and the false positives.} is 100\% reported to the set of probes which the ISPs validated.

\textbf{Negative Revelio Results.}
Out of the 5,121 lines Revelio tested, its results pointed to a simple NAT configuration (no CGN) for 3,454 {\em probes} in 63 different ISPs.
For the negative results, we obtained validation from 4 ISPs for which all \textit{probes} tested negative for upstream CGN in the ISP.
The 4 ISPs account for 508 {\em probes}.
We mention that (confirmed) negative results from Revelio testing do not preclude the existence of CGN technology in the corresponding networks.

Based on the ground truth we collected, we conclude that NAT Revelio did not generate any false negatives. 
Thus, provided that NAT Revelio can successfully run, its recall\footnote[5]{The recall represents the ratio between the number of true positives and the sum of the true positives and the false negatives.} is 100\% reported to the set of probes which the ISPs validated.
However, the Revelio methodology reported inconclusive results in 24\% of the cases (this numbers drops to 12\% if the measurement platform supports both UPnP and traceroute based tests). This is a consequence of Revelio's limitations, which we detail in Section~\ref{sec:limitations}.

\section{Analysis of Results}
\label{sec:analysis}
%%@@AD: separate the analysis section into pieces using paragraphs to add some structure. I think there are basically 3 different pieces of analysis here. state the takeaways from each analysis and why we did it.
The information we retrieved through the Revelio tests reveals additional insight about confirmed CGN operational setups, which we discuss next. 

\textbf{Number of hops between the CPE and the CGN}.
In 76 of the 92 positive cases, the CGN was located in the first IP hop in the ISP network. 
This is the case for all the 76 cases of the US satellite ISP that tested positive and as well for ISP\#6 (Zen). 
The other ISPs had 2 to 6 IP hops between the CPE and the CGN. 
This reflects two different CGN deployment architectures. 
The ISPs that exhibit only one hop between the CPE and the CGN deploy the CGN functionality in the first IP aggregation point in the network (e.g., the BRAS). 
This is consistent, for example, with some CGN cards that are available for insertion in the BRAS products. 
The other deployment setup installs the CGN in the ISP core network, allowing the aggregation of a higher number of costumers in a single CGN box.
This enables a more efficient multiplexing of the public IPv4 address pool.

\textbf{Addressing}. In terms of addressing used in the hops between the CPE and the CGN, ISP\#1 and ISP\#3 (Fastweb) assign shared address space~\cite{rfc6598} to devices in the access network. 
Interestingly, 
ISP\#1 uses shared address in the modem interface facing the home network and uses public IP address for all hops in the ISP network. 
ISP\#3 uses both shared addresses and private addresses (the first hop after the CPE uses shared address space and the subsequent ones private address space). 
The remaining ISPs use a mix of public and private addressing between the CPE and the CGNs.

%%@@AD: the GRA analysis is a bit different to the rest as it doesn't directly relate to CGN configuration. need to motivate this analysis
\textbf{GRA stability over time}. We analyzed the number of GRAs that we identified for each probe during the period we performed the tests. 
We found that the majority of the \textit{probes} (4,388 probes, accounting for 85\% of the total) have only one GRA during the whole tested period. 
The remaining 733 \textit{probes} use 2 to 19 different GRAs in the analyzed period. 
The average number of GRAs per \textit{probe} for all the \textit{probes} is 1.3 while the average number of GRAs per \textit{probe} for the \textit{probes} that tested positive for CGN is 3. 
In particular, for ISP\#1 the mean number of GRAs goes up to 4. 
We can see that while frequently changing the GRA is by no means exclusive of the CPEs behind a CGN, CGN deployments exhibits a considerably higher number of GRAs per customer.  
Additionally, we searched for GRAs simultaneously used by multiple \textit{probes}/CPEs. We found two GRAs simultaneously used in ISP\#1, one for 2 hours and the other for 48 hours. (Incidentally, we found 40 GRAs that were used by 2 \textit{probes} each, but we discovered that these cases were home networks that were hosting multiple Atlas probes. This is likely related to the incentives for hosting probes in Atlas).

%- additional insights
%GRA change frequency in CGNAT
%match the access link delay to the expected delay for a given access technology (in some cases there are different links in the home network that have one order of magnitude difference in the propagation delay, from a tens of micosecs to hundreds of microsecs, but still it is clear we didnt reach the access link. We can accomodate for these situations (and we did. We can tell how many false positives we fixed with this)
%possible elements that would disturb our methodology (homeplug, FEC, ICMP rate limiting) 

\section{Related Work}
\label{sec:related}

%Anecdotal evidence exists of broadband operators integrating CGN technology\cite{taxonomy_ton2014} in their growing networks to tackle IPv4 address space depletion.
The usage and impact of CGN-based solutions~\cite{taxonomy_ton2014} has lately drawn much attention from the community~\cite{rfc7021, rfc6269, bocchi2015impact}.
Consequently, we have seen several approaches for quantifying and confirming the degree in which operators are actively deploying operational CGNs~\cite{hotMB_cgn, richter2016multi}.    

Richter et al.~\cite{richter2016multi} use passive measurements to quantify CGN deployment rate in the Internet, after observing that some nodes in the BitTorrent DHT mistake addresses internal to a CGN for external addresses and therefore propagate these IP addresses to other nodes.
The authors report the average CGN penetration rate to be 17-18\% of all Eyeball ASes. 
In this paper, however, we focus on active CGN detection in fixed-line broadband providers at the IP level.
NAT Revelio empowers end-users to test whether their upstream provider connects them behind a CGN solution active in the access network.
Similar to our efforts, the Netalyzr~\cite{kreibich2010netalyzr} tool, initially meant as a networking debugging tool, has been repurposed to detect CGN solutions in mobile networks. 
The authors present the measurements results in~\cite{richter2016multi}, which corroborate the conclusion of prior work~\cite{triukose2012geolocating, Wang:2011:Untold} that for mobile providers CGN deployments are commonplace.
For wireline networks, however, it is non-trivial to detect and confirm CGN deployment on a per-IP basis.

Another notable effort towards passively quantifying the degree of CGN deployment is ~\cite{livadariu2017cgn}. 
The authors look at CGN deployment with a /24 IP prefix granularity.
For the 92 lines Revelio identified as being behind a CGN, we identified 52 covering /24 prefixes and cross-compared our results with ~\cite{livadariu2017cgn}.
We find that 38 prefixes appear as CGN-positive in ~\cite{livadariu2017cgn}, while 13 prefixes were classified as not being used in CGN deployments and one was unaccounted. 
Given that we were able to validate at the IP level 89 of these IPs, we thus highlight the benefits of an active measurements approach, while acknowledging the advantage of the passive approach to easily scale. 

Finally, using an earlier version of NAT Revelio, the authors presented in ~\cite{lutu2016nat} a smaller study on CGN deployment in the United Kingdom showing a very low penetration degree. This paper presents an evolution of the Revelio methodology that includes additional tests designed to support the myriad of technologies and devices deployed in real operational environments, as described in Section \ref{sec:methodology}.

%In~\cite{hotMB_cgn}, the authors propose \textit{NATAnalyzer}, an algorithm that combines multiple UDP packets, individual timeouts and traceroute measurements to actively discover previously unknown cascaded NAT middleboxes. 
%They evaluated NATAnalyzer it in a public field test combining mobile and wireline providers operating mainly in Germany and the US and found that majority of the ISPs already deploy CGN solutions. 

\section{Conclusions and Future Work}
\label{sec:conclusions}

Despite concerns about its performance impact, CGN solutions are part of the technology landscape during this ongoing phase of transition from IPv4 to IPv6. 
%Though anecdotal evidence for the presence of CGN technology in fixed-lines broadband providers exists, there have been limited efforts toward quantifying the presence of these ``Big NATs". 
Though passive data analysis allowed the quantification of CGN penetration in the current Internet ecosystem, we have no knowledge of a client-size only tool that can empower the end-user to determine the upstream NAT configuration. 
Revelio aims to fill this gap. 
Using NAT Revelio, we conducted a large-scale active measurement study for CGN detection targeting Europe and the U.S.
%More specifically, we deployed Revelio on two hardware-based measurement platforms: RIPE Atlas (with significant presence in Europe) and FCC's Measuring Broadband America (with significant presence in the United States). 
In total, we instrumented 5,121 measurement vantage points in over 60 different ISPs, capturing myriad combinations of home network topologies and Internet access technologies.
%Our results report limited use of the CGN technology in broadband providers.
We found that 10\% (6 out of 64) of the ISPs we tested have some form of CGN deployment.
We believe the combination of the FCC and RIPE Atlas study represents the largest active measurement study of CGN deployment in broadband networks to date.

We also validated our results at the probe level with representatives of 4 of the ISPs we tested (approx. 10\%).
Considering the ground truth we collected, we calculate the accuracy of our method at 100\%.
However, due to the limitation of the methodology, there are cases when some of the Revelio tests cannot run, thus hindering the test-suite efficiency. 
Overall, in 24\% of the lines we tested, Revelio gave inconclusive results. 
In the Atlas platform, where the \textit{probes} do not support UPnP, the rate of inconclusive results is 36\%.
This decreases significantly to only 12\% of the \textit{SK probes} in the FCC-MBA testbed, where Revelio was able to invoke UPnP actions.

For future work, we are planning on enhancing NAT Revelio with additional test to tackle its current limitations.
Additionally, we are exploring the possibility of adapting the test suite to deploy it on crowdsourcing platforms. 
Though this is not a trivial task, it would enable us to easily scale the number of measurement vantage points in the Internet and increase considerably the coverage of Revelio.

\bibliographystyle{acm}
\bibliography{./revelio}

\end{document}